# A MUSCLE MODEL BASED ON FELDMAN'S LAMBDA MODEL: 3D FINITE ELEMENT IMPLEMENTATION

Mohammad Ali Nazari[1,3], Pascal Perrier[1] and Yohan Payan[2]

(1) GIPSA-lab, CNRS UMR 5216
Grenoble Institute of Technology
Grenoble, France

(2) TIMC-IMAG, CNRS UMR 5525
Université Joseph Fourier
Grenoble, France

(3) Department of Mechanical Engineering
Faculty of Engineering, University of Tehran
Tehran, Iran

## INTRODUCTION

Stiffness properties of the human motor system depend on various physiological influences, such as passive elastic properties of muscle tissues, muscle activation, and neural feedback. Thus, muscle activation in motor systems not only induces changes in position, but also changes in stiffness. Stiffness changes and position changes intrinsically co-occur as the consequences of muscle activation, but to a certain extent they can also be controlled separately. Evidence supporting the hypothesis of these separate controls has been well documented in different studies that have shown the existence of (1) isometric motor tasks (change in muscle activation and stiffening, but no change in position), (2) isotonic motor tasks (change in position and in individual muscle activations, but without change in global muscle activation), and (3) unconstrained motor tasks (change in position and in muscle activation and stiffening) [1]. According to the literature, a specific control of stiffness is useful in motor control mainly for two reasons: increasing stiffness has an impact on movement speed and duration, and it is an efficient way to control movement accuracy, especially in the context of external perturbations. In general two types of functional motor or muscle models were provided in the literature: *adjustable stiffness* models and *adjustable starting length* models [2].

In biomechanics the reference muscle model is the Hill-type model [3], which is an example of adjustable stiffness model. The Feldman's lambda model is an adjustable starting length model [1]. A Hill-type model differs from the Feldman's lambda model from two main perspectives. First, in the Hill model the mechanical properties of the muscle (i.e. the relation between stress and strain) are based on measures when the muscle is maximally activated, while the Feldman's model is based on experimental data in unloading tasks. Second, the control variable is the force level for the Hill-type models, while force is the consequence of the specification of control variables for the Feldman's model. These control variables are defined in terms of threshold muscle lengths above which active force generation begins.

In this study, a 3D isometric element has been developed in the framework of a large deformation implicit finite element method. This element accounts for the constitutive law of an active muscle. The active part is embedded within the surrounding passive tissues that behave as a transversely isotropic hyperelastic material. This works has required adapting the Feldman's model to a distributed form, suitable in a 3D finite element. A comparison with a Hill-type muscle model is proposed.

## METHODS

The output force of a muscle ($F_m$) can be expressed as the resultant of the force in the passive elastic elements ($F_{PE}$) and the force generated in the contractile element ($F_{CE}$). Force in the contractile element is a function of muscle length ($L$), muscle velocity ($v$) and activation ($A$).

$$F_{CE}=f(L,v,A) \quad (1)$$

This force is usually expressed in a multiplicative way as a product of three distinct functions: force-length ($F_L$), force-velocity ($F_v$) and time transition function of the activation ($f_A$) [3]:

$$F_{CE}= f_A \times F_L \times F_v \quad (2)$$

This multiplicative account of contractile force is called adjustable stiffness because the contractile element behaves like a nonlinear spring which stiffness varies as a linear function of force for different activation levels [2]. The Hill-type models are categorized in this group.

In another approach, the contractile force is expressed as a general nonlinear function of activation level and velocity:

$$F_{CE}=f(L,v,l_{threshold})\quad(3)$$

In this subgroup, the activation ($A$) depends on the difference between the muscle length and the zero-force length, which is the control parameter, like in Feldman's model [1]. These models are called adjustable starting length models [2]. In these models, muscle stiffness changes as a nonlinear function of force for different activation levels. In Feldman's model, for a given activation command, muscle force is the consequence of the combination of the stretch reflex mechanism and the force-length characteristics, which isan exponential curve, called the invariant characteristic (IC), described as [4]:

$$F_{CE\_Feldman}=F_{max}(\exp([l(t-d)-l_{threshold}+\mu v(t-d)]^{+}/l_c)-1)\quad(4)$$

where $F_{max}$ is the maximum force generation capacity of a muscle and is a function of the physical cross sectional area (PCSA) of the muscle, $l_{threshold}$ is zero-force length or threshold length, $l_c$ is a characteristic length, $v$ is muscle velocity and $\mu$ is a damping coefficient. Both muscle length and velocity in this equation are delayed values at time $t$-$d$. $[]^+$ means that the force is equal to zero if the expression within $[]^+$ is negative. A passive force should be added to this active force to take into account the passive mechanical property of the muscle.

Extension of a global muscle model to a continuum needs the design of a distributed version of that model. In a distributed model all lumped quantities are replaced with their distributions: force terms are replaced with Cauchy stresses and length quantities are replaced with stretch ratios. Starting from equation (4), the active Cauchy stress in the so-called Distributed Feldman's Model (DFM) becomes:

$$\sigma_{CE\_Feldman}=\sigma_{max}(A_{pcsa}/A)(\exp([\lambda(t-d)-\lambda_{threshold}+\mu v(t-d)]^{+}(l_0/l_c))-1)\quad(5)$$

In this relation $\sigma_{max}=F_{max}/A_{pcsa}$ is the maximum stress generation capacity of the muscle.

In the finite element implementation of a muscle model, the active part works in parallel to the surrounding passive tissues as proposed in [5]. So the resultant stress of the muscle in the muscle's fiber direction is the sum of the active and passive stress:

$$\sigma_m=\sigma_{CE}+\sigma_{PE}\quad(6)$$

Surrounding tissues behave like a nearly incompressible transversely isotropic hyperelastic material. This is characterized with a strain energy function.

## RESULTS

The introduction of Feldman's muscle model in a three dimensional continuum finite element model has shed light on its behavior which otherwise was not evident in its original formulation. Its behavior shows that addition of a passive surrounding tissue shifts the threshold length (Figure 1). This also is in agreement with the results mentioned in [2]. This shift in the values of $\lambda<1$ is toward right (which means less active force for the same muscle length) and for $\lambda>1$ is toward left. Hence, in calculating the equilibrium point in an agonist-antagonist pair, the effect on one muscle is cancelled by the effect on the other muscle and the model gives the same result as the original one dimensional model.

The comparison between a Hill-type muscle model and DFM in generating a voluntary action pseudostatically doesn't show much differences (Figure 2).

According to the experiments and as is the case in a Hill-type model the muscle force cannot pass the maximum voluntary action, hence the DFM should be combined with this limit such as the muscle force does not pass maximum voluntary action force (last upper curve of hill-type curves in Fig. 2).

Both models are used in a 3D Biomechanical finite element face model [6] in the study of lip protrusion due to the activation of orbicularis oris muscle (Figure 3). Their effect does now show much difference in the final shape. The only difference comes with the time history evolution of this motion.

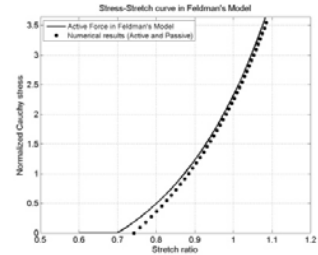

**Figure 1.** Comparison of the Stress-Strain curve in the Feldman's model in absence (solid line) *versus* in presence (dotted line) of passive properties.

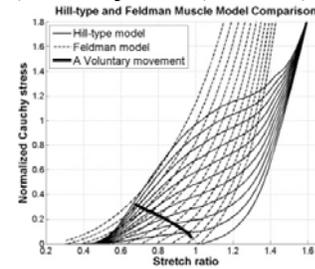

**Figure 2.** Comparison of a Hill-type muscle model and Feldman's model. Differences are small below the rest length (stretch ration=1), which is usually the case in contracted muscles. However it can be important above the rest length.

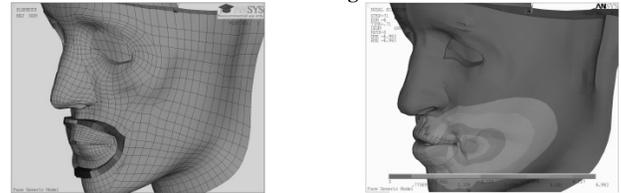

**Figure 3.** (Left) Orbicularis oris muscle (Right) final shape due to its activation by both muscle models.

## DISCUSSION

The presentation of a muscle model based on the results of the unloading experiment provides a simple way to simulate such behavior which would be accounted for by a complex activation dynamics in a Hill-type model. The Feldman's lambda model is a well-known model in the domain of motor control studies but is rarely utilized in the biomechanical studies. Its introduction in a finite element model provides a physical meaning for its parameters. It provides also ideas for the design of new possible experiments to characterize these parameters. The integration of both a Hill-type model and the Feldman's model in a 3D biomechanical face model has provided a useful basis for studying their impact in real application [6]. Among these applications study of speech facial gestures is in progress.